\documentclass[]{aa}
\usepackage{psfig}
\usepackage{txfonts}
\usepackage{graphicx}
\usepackage{mathrsfs}

\begin{document}

\title{The relationship between the prestellar core mass function 
and the stellar initial mass function}
\author{Simon\,P.\,Goodwin$^1$, D.\,Nutter$^2$, P.\,Kroupa$^3$, 
D.\,Ward-Thompson$^2$ \and A.\,P.\,Whitworth$^2$}

\authorrunning{S.\,P.\,Goodwin et al.}
\titlerunning{The relationship between core and stellar IMFs}

\offprints{S.Goodwin@sheffield.ac.uk}

\institute{$^1$Department of Physics \& Astronomy, University of
  Sheffield, Hicks Building, Hounsfield Road, Sheffield, S3 7RH, UK.\\
$^2$School of Physics \& Astronomy, Cardiff University, Queens Buildings,
The Parade, Cardiff, CF24 3AA, UK. \\
$^3$Argelander-Institut f\"ur Astronomie, Universit\"at Bonn, Auf dem
  H\"ugel 71, D-53121 Bonn, Germany.}

%\date{7/5/03}

\abstract{Stars form from dense molecular cores, and the mass function
  of these cores (the CMF) is often found to be similar to the form of
  the stellar initial mass function (IMF).  This suggests that the
  form of the IMF is the result of the form of the CMF.  However, most
  stars are thought to form in binary and multiple systems, therefore
  the  relationship between the IMF and the CMF cannot be trivial.  We
  test two star formation scenarios - one in which all stars form as
  binary or triple systems, and one in which low-mass stars form in a
  predominantly single mode.  We show that from a log-normal CMF,
  similar to those observed, and expected on theoretical grounds, the
  model in which all stars form as multiples gives a better fit to the
  IMF.

 }

\maketitle

\section{Introduction}

The origin of the stellar initial mass function (IMF) is one of the
outstanding unsolved problems in astrophysics.  As stars form in dense
molecular cores (see e.g. Ward-Thompson et al. 1994; Kirk et al. 2005;
Ward-Thompson et al. 2007), it might well be expected that the IMF
is related to the mass function of those cores (the CMF).  This
idea is supported by observations of prestellar cores, which show that
their mass functions are often similar to the IMF of  Galactic
field stars (Motte et al. 1998; Testi \& Sargent 1998; Johnstone et
al. 2000; Johnstone et al. 2001; Motte et al.  2001; Johnstone \&
Bally 2006; Alves et al. 2007; Young et al. 2006;  Nutter \&
Ward-Thompson 2007; Simpson et al. 2007).  Further support is given by
the observation that Taurus may have both an unusual CMF (Onishi et
al. 2002) and an unusual IMF (Luhman 2004; see also Goodwin et
al. 2004c), although Kroupa et al. (2003) show that the IMF in Taurus may be
compatible with the field IMF.

However, the relationship between the CMF and the IMF cannot be
simple, as many, if not the vast majority, of stars form in binaries or
higher-order multiple systems (see Goodwin \& Kroupa 2005; Duch\^ene
et al. 2007 and Goodwin et al. 2007 and references therein; see also
Clark et al. 2007).  Observations suggest that the binary frequency amongst
young stars is higher than in the field (see Goodwin et al. 2007 and
references therein) implying that binaries are destroyed by dynamical
interactions in clusters (see Kroupa 1995a,b).  However, Lada (2006)
has argued that most M-dwarfs  form as single stars, since the {\em field}
M-dwarf binary fraction is  relatively low and there is no need to
invoke dynamical destruction of low-mass binaries to form these
(single) stars.  The opposing view is argued by Goodwin \& Kroupa
(2005) and  Goodwin \& Whitworth (2007).

If stars (or at least relatively high-mass stars) usually form in
small-$N$ multiples then there cannot be a trivial one-to-one
relationship between the IMF to the CMF.  Firstly, the mass of a  core
is distributed between a number of stars. Secondly, some  stars are
expected to be ejected at an early age from small-$N$ multiples
(e.g. Reipurth \& Clarke 2001; Goodwin et al. 2007 and references
therein; see also Section~3).  Thirdly, many binary systems are
expected to be destroyed in  clusters (Kroupa 1995a,b; Kroupa et
al. 2003; Goodwin \&  Whitworth 2007; also see Goodwin et al. 2007 and
references  therein).  Thus the CMF should relate most closely to the initial
{\em system} mass function which, in turn, is modified by dynamical
effects to produce a mixture of single and multiple systems.  

In this paper we examine the relationship between the IMF and the CMF,
in particular we use the new results for the CMF in Orion from Nutter
\& Ward-Thompson (2007).  In Section 2 we review observations of the
CMF, in Section 3 we present our general method, and in Section 4 we
compare the IMFs we produce with the observations.

%%%%%%%%%%%%%%%%%%%%%%%%%%%%%%%%%%%%%%%%%%%%%%%%%%%%%%%%%%%%%%%%%%%%%%%%%%
\section{Observations of the CMF}

The first observational link between the IMF and the CMF was made by
Motte et al. (1998) in a millimetre study of the $\rho$-Ophiuchi
molecular cloud. They found that the high-mass slope of the CMF
matched that of the IMF. This result has been confirmed for Ophiuchus
(Johnstone et al. 2000; Young et al. 2006; Simpson et al. 2007) and a
number of other nearby clouds, including Orion (Motte et al. 2001;
Johnstone et al. 2001; Johnstone \& Bally 2006; Nutter \&
Ward-Thompson 2007), the Pipe Nebula (Alves et al. 2007), 
and Taurus (Onishi et al. 2002; however see
Goodwin et al. 2004c), as well as for more distant massive
star-forming regions such as NGC 7538 and M17 (Reid \& Wilson 
2006a,b).

While the slope of the CMF seems to be consistent from region to
region, the position of the peak of the CMF appears to shift from
$\sim\! 0.1~M_\odot$ in nearby low-mass regions such as $\rho$-Ophiuchus
(e.g. Motte et al. 1998), to a higher mass of $\sim\! 1~M_\odot$ in more
distant and massive star-forming regions such as Orion (e.g. Nutter \&
Ward-Thompson 2007). Very massive star-forming regions such as M17
show a flattening of the CMF at an even higher mass of $\sim\!
8~M_\odot$ (Reid \& Wilson 2006a,b), though the data are incomplete
before a turn-over is seen. Whether this is an intrinsic effect where
the mass of the peak in the CMF is related to the mass of the stars
being formed, or an observational effect caused by the blending of
multiple sources at larger distances, is not yet known.

%%%%%%%%%%%%%%%%%%%%%%%%%%%%%%%%%%%%%%%%%%%%%%%%%%%%%%%%%%%%%%%%%%%%%%%%%%
\section{The relationship between core and stellar mass functions}

We assume that the star formation properties of a core may be
described by three basic parameters: the mass of the core $M_C$, the
efficiency with which the core turns gas into star(s) $\epsilon$ (so
that the total mass of stars is $\epsilon M_C$), and the number of
stars formed within a core $N_*$ (the choices of $N_*$ are 
discussed in section~4).  We note that $\epsilon$ and $N_*$
may well be functions of $M_C$.

There are then two basic distribution functions of the stars.   The
multiple system mass function (MSMF) is the mass distribution of the
multiple systems produced by cores.  The single star mass function
(SSMF) is the mass distribution of all of the {\em individual} stars
formed in all of the single and multiple systems in all of the 
cores.  It is the
SSMF that will correspond to the (binary corrected) initial mass
function (IMF; see below).

Note that the MSMF will evolve due to several effects.  Firstly,
unstable high-order multiple systems may decay, preferentially
ejecting single, low-mass stars (e.g. Reipurth \& Clarke 2001; Bate
et al. 2002, 2003; Sterzik \& Durisen 2003; Goodwin et al. 2004a,b;
Delgado Donate et al. 2004a,b;  Hubber \& Whitworth 2005; Umbriet et
al. 2005).  Secondly, binaries  may be `destroyed' by rather violent
close  binary-binary/binary-single interactions (Kroupa 1995a,b), and
thirdly, wide, low-mass binaries may be `disrupted' by  more gentle
impulses from passing stars (Goodwin \& Whitworth  2007).

\begin{figure*}
\centerline{\psfig{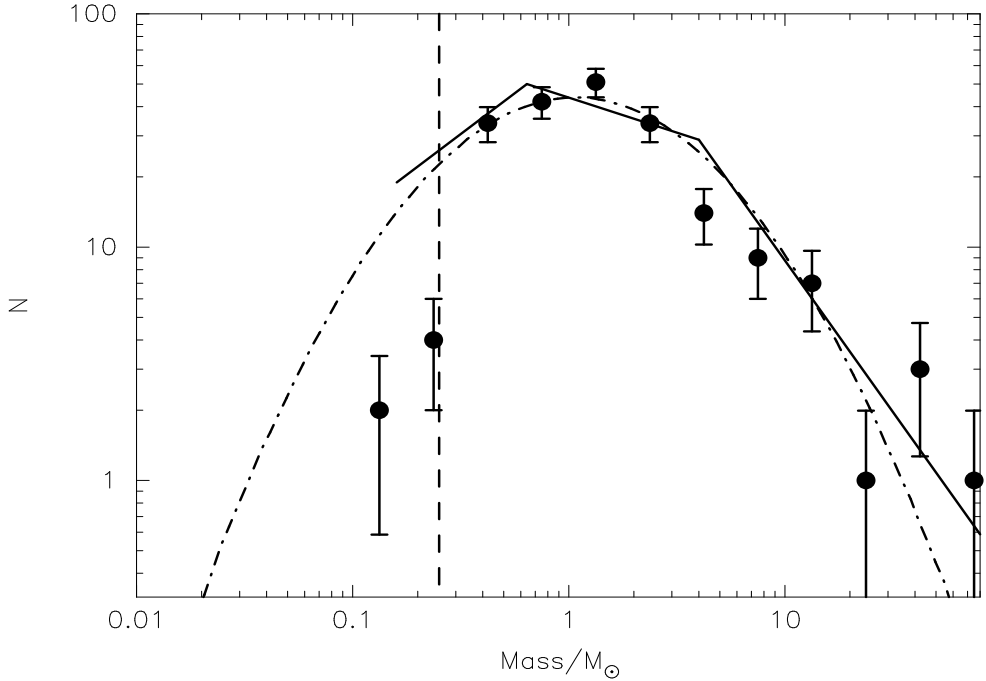}}
\caption{The CMF of Nutter \& Ward-Thompson (2007) with $\sqrt{N}$ error-bars,
  the completeness limit of the observations is marked by the vertical
  dashed line.  The data are
  fitted by a log-normal distribution with mean $\mu_{{\rm log}_{10}
  M} = 0.05$ and $\sigma_{{\rm log}_{10} M} = 0.55$ (dashed-dot 
  line).  Also plotted for
  comparison is the canonical IMF (eqn.~\ref{eqn:imf}) with the same
  slopes ($\alpha_1 = 0.3$, $\alpha_1 = 1.3$ and $\alpha_1 = 2.3$),
  but turn-over masses increased by a factor of 8 ($M_0 = 0.16 M_\odot$, 
$M_1 = 0.64 M_\odot$, $M_2 = 4 M_\odot$ and $M_3 = 80 M_\odot$).}
\label{fig:cmffit}
\end{figure*}

\subsection{From a CMF to an IMF}

The procedure for generating an IMF (via a MSMF and SSMF) from  a CMF
is very simple.  We randomly sample a core mass from a CMF (see
Section~3.3 for our choice of CMF).  This core then produces $N_*$
stars of {\em total} mass $\epsilon M_C$ (this is similar to the
approach of Sterzik et al. 2001, however we do not constrain the
IMF of the stars in any way except through the CMF of the cores).

The masses of the $N_*$ components in a multiple system are chosen
randomly.  In a binary system ($N_*=2$), the masses of the primary and
secondary are selected from a flat mass ratio distribution (ie. one
random number U$[0,1]$).  In a higher-order multiple the masses are 
distributed randomly, ie.  $N_*$ random numbers U$[0,1]$ are chosen
and then the sum is  normalised to unity to provide the mass
distribution.

The SFE, $\epsilon$, is chosen to provide the best fit to the canonical
IMF (see below) and is assumed to be constant for all cores.  It might
be thought that SFE should depend on  the mass of stars formed,
as feedback energy increases with increasing stellar mass.  However,
the potential well from which gas must be removed by feedback also
increases with increasing stellar and gas mass and so possibly
the SFE is constant, or even increasing with mass.  Given the
uncertainties involved we make the simplest assumption possible that
the SFE is constant.  As will be seen, a good fit to the IMF can be
obtained while making this assumption and there appears no need to
appeal to a variation of SFE between cores of different masses.

\subsection{The canonical observed IMF}

We assume that the actual underlying IMF of stars has the 
canonical form (Kroupa 2002; see also Kroupa 2007)

%\begin{displaymath}
\begin{equation}
N(M) \propto \left\{
\begin{array}{ll}
M^{-\alpha_1} & \,\,\,\, M_0 \leq M \leq M_1 \\
\left( \frac{M_1^{-\alpha_1}}{M_1^{-\alpha_2}} \right) M^{-\alpha_2} & \,\,\,\, M_1 \leq M \leq M_2 \\ 
\left( \frac{M_1^{-\alpha_1}}{M_1^{-\alpha_2}} \right) \left(
\frac{M_2^{-\alpha_2}}{M_2^{-\alpha_3}} \right) M^{-\alpha_3} & \,\,\,\, M_2 \leq M \leq M_3 \\
\end{array} \right.
\label{eqn:imf}
\end{equation}
%\end{displaymath}

\noindent with $\alpha_1 = 0.3$, $\alpha_2 = 1.3$ and $\alpha_3 = 2.3$ 
as the slopes, and $M_0 = 0.02 M_\odot$, $M_1 = 0.08 M_\odot$, 
$M_2 = 0.5 M_\odot$ and $M_3 = 10 M_\odot$ as the masses of the 
limits and turning points of the IMF.  This form of the IMF matches
well other recent determinations of the IMF (e.g. Chabrier 2003).

The canonical IMF is corrected for the presence of unresolved binary
systems and therefore the IMF should follow the SSMF.

\subsection{Forms of the CMF}

Our standard CMF is that determined for Orion by Nutter \& 
Ward-Thompson (2007).  We model this as a log-normal with dispersion
$\sigma_{{\rm log}_{10} M} = 0.55$ and a mean of $\mu_{{\rm log}_{10}
  M} = 0.05$ as illustrated
in Fig.~\ref{fig:cmffit}.  This CMF is not too
dissimilar to an IMF-like distribution with the slopes $\alpha$
remaining the same, but turn-over masses of $\sim\! 8$ times those in
the canonical IMF (eqn.~\ref{eqn:imf}).  We also note that this is very
similar to the CMF of the Pipe dark cloud (Alves et al. 2007).

Our fit to the Nutter \& Ward-Thompson (2007) CMF is above
the final two points which are below the completeness limit.  We
assume that the {\em entire} core mass distribution is modelled by a
log-normal which would be expected on theoretical grounds (Padoan \& Nordlund
2002,2004; Klessen \& Burkert 2000; Klessen 2001; Li et al. 2003;
Jappsen et al. 2005) even in a wide variety of physical conditions
(see esp. Padoan \& Nordlund 2002; Jappsen et al. 2005).  We also note
that Padoan \& Nordlund (2002,2004) tend to find a very steep decline 
in the CMF below the peak which would be compatible with these
observations. 

Clark et al. (2007) recently noted that the free-fall times of clumps
of different masses are different and that low-mass cores collapse
significantly faster than higher-mass cores.  Thus an observed CMF
must be constantly replenished with low-mass cores in order to retain
a constant form.  We assume that our CMF represents a `snapshot' of
the CMF.  If star formation occurs rapidly in clusters (in $<1$~Myr,
e.g. Elmegreen 2000) then the observed CMF should represent the total
CMF for masses above the peak of the CMF as the free-fall time for
cores $> 1 M_\odot$ is a significant fraction of the cluster formation
timescale (Clark et al.  2007, see their fig.~1).

%%%%%%%%%%%%%%%%%%%%%%%%%%%%%%%%%%%%%%%%%%%%%%%%%%%%%%%%%%%%%%%%%%%%%%%%%%%
\section{Results}

\subsection{The fully multiple model}

Goodwin \& Kroupa (2005) suggested that the  observed properties of
multiple systems could be reproduced if  each core produces 2 or 3
stars. Single field  stars are then produced by the dynamical decay
and destruction of multiple systems in young clusters (Kroupa 1995a,b;
Goodwin \& Kroupa 2005; Goodwin \& Whitworth 2007; Goodwin et al. 2007
and references therein).

To model the fully multiple scenario we assume that cores of mass
$\epsilon M_C < 0.75 M_\odot$ form entirely binary systems, and
cores with $\epsilon M_C \ge 0.75 M_\odot$ form multiple systems with 
a ratio of $3\!:\!1$ binaries-to-triples.  The SFE is chosen to give 
a good fit to the canonical IMF with $\epsilon = 0.27$.  

In this scenario the {\em initial} multiplicity fraction is unity.
Single stars and brown dwarfs are produced by the destruction of 
many (especially low-mass) initially multiple systems (see 
Section~3).  

\begin{figure*}
\centerline{\psfig{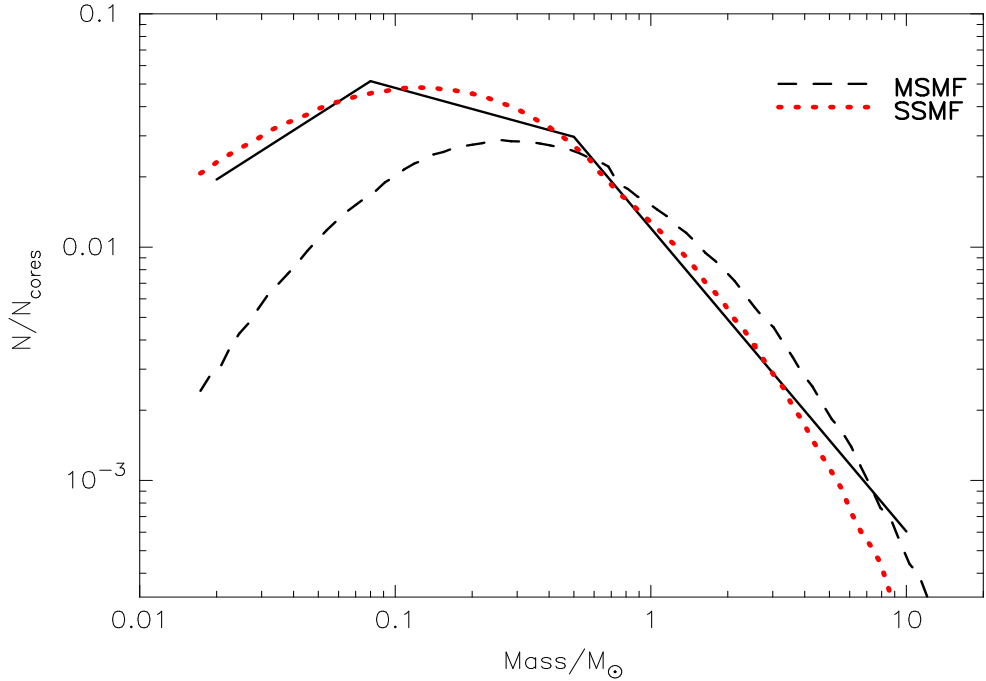}}
\caption{The fully multiple model with $N_\star = 2$ for systems with
  $\epsilon M_C < 0.75 M_\odot$ and $N_\star = 3\!:\!1$
  binary-to-triple ratio for $\epsilon M_C \ge 0.75 M_\odot$,
  and $\epsilon = 0.27$.  The initial MSMF (dashed line), and the 
  SSMF (red dotted line) compared to the canonical IMF (thin solid lines).}
\label{fig:gk}
\end{figure*}

The result of the fully multiple model are illustrated in Fig.~\ref{fig:gk}.  This
model produces a good fit to the canonical IMF for all masses except
the very highest.  The mass
functions dip below the canonical slope of $-1.3$ at high masses due
to the steep decline of the log-normal CMF at high masses\footnote{It might
be expected that the actual CMF continues as a power-law decline 
rather than being fitted by a log-normal at high masses (see
e.g. Padoan \& Nordlund 2002,2004).}  

%The good fit to the canonical IMF is obtained because the
%initial log-normal shape of the CMF is retained as fragmentation is
%almost independent of core mass.  The observed IMF is then skewed towards
%lower masses by the preferential destruction of low-mass systems.

The fully multiple model requires the dynamical destruction (see
e.g. Kroupa 1995b; Kroupa et al. 2003; Goodwin \& Kroupa 2005; 
Goodwin et al. 2007) of significant numbers of low-mass binary systems
in young clusters in order to change the initial binary fraction of
unity to the field value.

We note that in this model brown dwarfs are not primarily produced as
single objects in cores (`star-like' formation, e.g. Padoan \&
Nordlund 2004), nor as ejected embryos from high-mass cores (the
ejection hypothesis, e.g. Reipurth \& Clarke 2001).  Instead they
mainly form as the distant companions to M-dwarfs which are then
disrupted. This is the scenario proposed by Goodwin \& Whitworth
(2007) as a major mode of brown dwarf formation.  We note that this
might be consistent with the idea that brown dwarfs form as a separate
population of objects, possibly with a discontinuous IMF (Kroupa et
al. 2003; Thies \& Kroupa 2007; Kumar \& Schmeja 2007).

\subsection{The Low-mass single star model}

Lada (2006) suggests that M-dwarfs tend to form as single stars, since
most M-dwarfs in the field (roughly 55\% by total number) are single
stars.  In this picture, destructive dynamical processes are 
unimportant for low-mass stars.  However, Solar-type stars (and 
higher-masses) must still usually form
in multiple systems to fit the observed high multiplicity fraction of
T Tauri stars (e.g. Mathieu 1994; Patience et al. 2002; Duch\^ene et
al. 2007 and Goodwin et al. 2007 and references therein).

We model the situation where most low-mass stars form as singles (the
low-mass single star model) by making $N_*$ a strong
function of the core mass.  Cores with $\epsilon M_C < 0.5
M_\odot$ form stars in a $2\!:\!1$ single-to-binary ratio.  Cores with 
masses $0.5 < \epsilon M_C/M_\odot  < 1 M_\odot$ have a $1\!:\!1$ ratio of
singles-to-binaries. And cores with  $\epsilon M_C \ge 1 M_\odot$
have a $3\!:\!1$ binary-to-triple ratio (as in the fully multiple 
model).  These probabilities roughly
reflect a combination of the low-mass field (M-dwarf, see Fischer \&
Marcey 1992, also Lada 2006) and intermediate-mass PMS binary
fractions (see Duch\^ene et al. 2007 and Goodwin et al. 2007 and
references therein).

\begin{figure*}
\centerline{\psfig{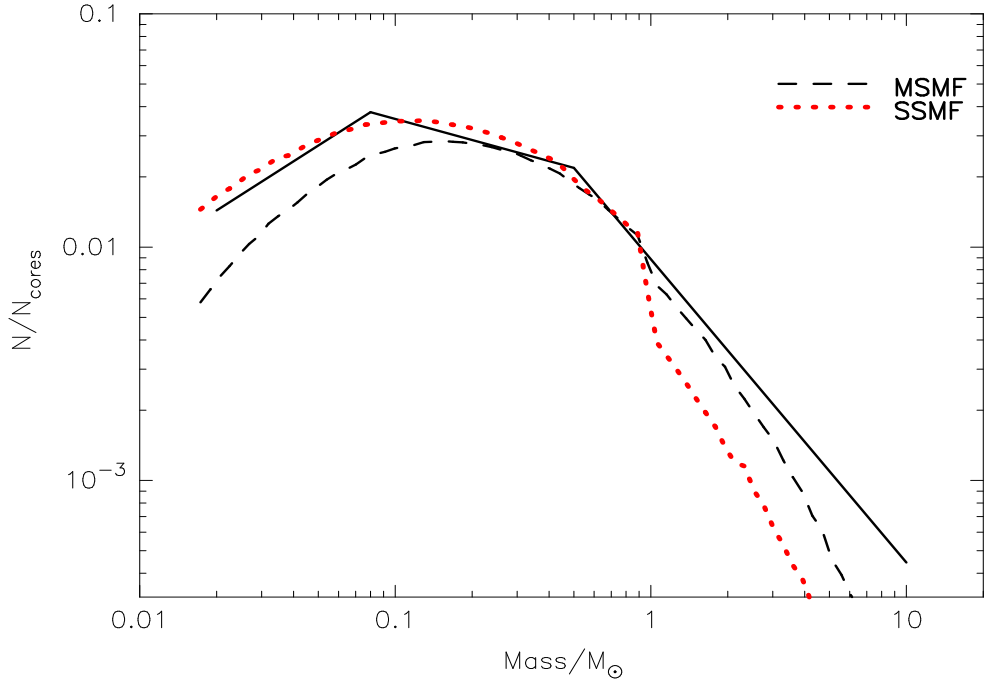}}
\caption{Low-mass single star model with $N_\star =$ $2\!:\!1$ single:binary
  for $\epsilon M_C < 0.5 M_\odot$, $1\!:\!1$ single:binary
  for $0.5 < \epsilon M_C/M_\odot < 1$, and $3\!:\!1$ binary:triple for
  $\epsilon M_C \ge 1 M_\odot$, with $\epsilon = 0.15$.  The initial MSMF 
  (dashed line), and the SSMF (red dotted line) compared to the 
  canonical IMF (thin solid lines).}
\label{fig:lada}
\end{figure*}

The results from this model are shown in Fig.~\ref{fig:lada} with a
SFE of $\epsilon =  0.15$ which provides the best (but still not a
good) fit to the canonical IMF.  This model cannot be made to fit 
the canonical IMF well with any choice of $\epsilon$.  

The main problem with the low-mass single star model is that it provides no mechanism
for forming brown dwarfs other than as single brown dwarfs in very low
mass cores.  As the CMF peaks at $\sim\! 1 M_\odot$ there are very few
cores below $0.1 M_\odot$ and so the SFE must be made very low in
order to shift the peak of the CMF to the peak of the IMF.  This is
avoided in the fully multiple model by producing brown dwarfs as companions to
low-mass stars.  Increasing the SFE to greater than 15\% results in
far too few VLMOs and also shifts the peak of the IMF to too-high a
mass.  We also note that a SFE of $15$\% suggests that a significant
number of cores produce planetary-mass objects (which are not shown in
Fig.~\ref{fig:lada}). 

There are two ways we might possibly escape from the problem of
too-few brown dwarfs.  Firstly, we may postulate a very
large population of low-mass cores from which brown dwarfs can form
(e.g. Padoan \& Nordlund 2004).  In this situation the vast majority
of cores would be below the peak of the CMF.  However, the simulations 
of Padoan \& Nordlund (2002,2004) suggest that the CMF should drop
sharply below the peak, far faster than a log-normal distribution.
Thus, there appears to be no good reason for expecting a large
population of low-mass cores (see Goodwin \& Whitworth 2007 for a
number of other reasons why significant numbers of low-mass brown
dwarf-forming cores would not be expected).

Secondly, we may form brown dwarfs by ejecting low-mass embryos from
massive cores, thus creating brown dwarfs (e.g. Reipurth \& Clark
2001).  However, large numbers of ejections per high-mass core  would
be required to create almost all of the brown dwarf population in this
way, and large numbers of such ejections have consequences that are
difficult to reconcile with observations (see e.g. Goodwin \& Kroupa
2005; Goodwin et al. 2007 and Whitworth et
al. 2007 and references therein; Goodwin \& Whitworth 2007).

The slope of the high-mass end of the IMF is also far too 
steep ($\sim\! 1.6$
rather than the observed $\sim\! 1.3$).  This is due to the change in
the modes of fragmentation at $\epsilon M_C = 0.5 M_\odot$ and $1
M_\odot$\footnote{Small jumps may be seen at $0.5 M_\odot$ and $1
M_\odot$ as the function steps suddenly between  different modes
rather than being smooth, however
a smooth function merely evens-out these jumps but does not change the
overall appearance of the IMF.}.  This is because at high masses the
mass of stars is divided between 2 or 3 stars, while at lower masses
it is divided only between 1 or 2.  Thus altering the modes of
fragmentation changes the slope from the IMF-like slope of the CMF to a
steeper slope.  The upper-mass slope only matches the slope of the CMF
if fragmentation is independent of mass for cores above the knee in
the IMF at $0.5 M_\odot$. 

The problem with the upper-mass slope can be alleviated somewhat by
assuming that all stars form with the field binary fraction (rising
from 33\% for M-dwarfs to 60\% for G-dwarfs).  However, this solution
conflicts with observations that the initial binary fraction for stars
$>1 M_\odot$ is consistent with unity (Goodwin \& Kroupa 2005;
Duch\^ene et al. 2007; Goodwin et al. 2007).  These observations
suggest that there must be a fairly rapid transition between $0.5$ and
$1 M_\odot$ from a low to a high primordial binary fraction which will
result in too-steep an upper-mass slope of the IMF.

We have assumed that the mass ratio of binaries is a flat
distribution.  Biasing the mass ratio distribution to low-$q$
(ie. highly unequal mass systems) improves the problems at the
high-mass end of the IMF slightly.  If most high-mass cores produce
one large star and one or two very low-mass stars, then the IMF at the
high-mass end becomes more similar to the MSMF (as this is dominated
by one of the stars).  However, the mass ratio distribution needs to
be very biased for this to have a significant effect.

The too-steep slope of the upper-end of the IMF can also be solved by
assuming that the SFE increases with increasing core mass (in just the
right way). However, we feel this solution is unlikely as the SFE
would have to be fine-tuned to give the correct slope and it would
seem peculiar to postulate that low-mass cores produce stars at very
low efficiencies ($\sim 10$\%), whilst higher-mass cores are 
able to convert more of their gas ($\sim 30$\%) into stars (the 
opposite of what might be expected from arguments based on feedback).

%%%%%%%%%%%%%%%%%%%%%%%%%%%%%%%%%%%%%%%%%%%%%%%%%%%%%%%%%%%%%%%%%%%%%%%%%%
\section{Discussion}

We have shown that the observed mass functions of cores in Orion
 B (Nutter \& Ward-Thompson 2007) can give rise to the IMF of 
stars.  In particular we have shown that, 
to produce the stellar {\em and sub-stellar} IMF, the majority of 
these cores must fragment into multiple systems.  However, there are a
number of issues about cores and the CMF that are worth discussing in
this context.

It should be noted that it may not be fragmentation into 
`cores' in clusters that sets the IMF of stars.  If competitive
accretion (see Bonnell et al. 2007 and references therein) is 
the dominant process, then the CMF at best acts to set the initial 
masses upon which competitive accretion begins to work.  In such a
scenario there would be little or no relationship between the CMF and 
the IMF.

However, we would argue that the form of the CMF in diffuse star
forming regions {\em does} have a direct relevance to the origin and
form of the IMF.  Given the apparent universality of the IMF across a
wide range of star forming environments (e.g. Kroupa 2002)
we are presented with two options.  Firstly, that the mechanism(s)
that produce the IMF are fundamentally different in different
environments, but they always produce the same outcome.  Or, secondly,
that there is a single, underlying, mechanism that produces the IMF in
all environments.  The latter possibility appeals due to its
simplicity, and would suggest that the form of the CMF is the driving
factor in establishing the form of the IMF, and that the form of the
CMF is roughly the same in diffuse and clustered regions (even if the
cores themselves are different in spacial size).  Indeed, simulations
of turbulence always seem to produce roughly log-normal CMFs whatever
the environment.

%%%%%%%%%%%%%%%%%%%%%%%%%%%%%%%%%%%%%%%%%%%%%%%%%%%%%%%%%%%%%%%%%%%%%%%%%%
\section{Conclusions}

We have examined the relationship between the core mass function
(CMF) and the stellar initial mass function (IMF).  We use the Orion 
CMF from Nutter \& Ward-Thompson (2007) as a `standard' which
we fit using a log-normal distribution.  We note that this CMF
is not dissimilar to the stellar (Kroupa 2002) IMF shifted upwards in
mass by a factor of $8$ (see also Alves 2007).  We randomly sample
cores from the CMF and assumed that each core produces
a certain number of stars with a random distribution of masses between
the components.

The canonical IMF is reproduced very well by a scenario in which
every low-mass cores fragment into binaries, and high-mass cores 
fragment into a multiple system with a ratio of
binaries-to-triples of $3\!:\!1$ (see e.g. Goodwin \& Kroupa 2005) and 
a star formation efficiency (SFE) of $\sim\! 30$\%.  Dynamical 
disruption (Kroupa 1995a,b; Goodwin \& Whitworth 
2007) of systems then evolves the initial binary fraction of unity 
into the field population.

We find that a scenario in which low-mass stars preferentially form
single systems (e.g. Lada 2006) cannot reproduce the observed IMF from
a log-normal CMF.  Firstly, the slope of the high-mass IMF is too
steep.  Secondly, and most seriously, this model cannot reproduce the
correct numbers of brown dwarfs to high-mass stars.  The best-fit to
the canonical IMF is found when the SFE is only $\sim\! 15$\%.  Such a
low SFE is required, as the only way in which brown dwarfs may be
produced in significant numbers is through the formation of a single
brown dwarf from a core.  Higher SFEs are required to produce
sufficient high-mass stars, however such SFEs significantly
under-produce brown dwarfs and low-mass stars.

A lingering question is the value of the star formation efficiency
that must be applied to fit the IMF.  The best-fit value of
$\epsilon$ in the fully multiple model suggest that 
only $\sim\! 30$\% of the mass in a core ends-up
in the stars which that core forms (a similar value for the SFE is
found by Alves et al. 2007).  This seems a very low value and
may suggest that the determinations of the absolute core masses
are wrong.  Another
possibility is that feedback from jets is far more efficient than
previously thought and manages to disperse most of the gas initially
in the core.  A final possibility is that we are not observing
`typical' cores which produce the IMF and that the observed CMFs will produce
somewhat top-heavy IMFs (cf. Taurus, Goodwin et al. 2004c).

We conclude that a model in which {\em all} stars and brown 
dwarfs form in multiple systems from a log-normal core mass distribution 
provides a very good fit the observed IMF. 

\begin{acknowledgements}

SPG was supported during some of this work by a UKAFF Fellowship.  SPG also
acknowledges the support and hospitality of the International Space
Science Institute in Bern, Switzerland where part of this work was
done as part of a International Team Programme. SPG and PK also
acknowledge support from a Royal Society International Joint Project 
grant.  PK also 
thanks the Leverhulme Trust for the award of a Leverhulme Trust Visiting
Professorship at Sheffield University.

\end{acknowledgements}

\end{document}